\documentclass[a4paper,12pt]{article}
\usepackage{graphicx,bbm}
\usepackage{amsmath,amsfonts,amssymb}
\usepackage{fleqn} 
\usepackage{mathrsfs}
\usepackage{epsfig}\parskip 5pt plus 1pt
\usepackage[footnotesize]{caption}

\def\I{\mathrm{i}}

\def\beq{\begin{equation}}
\def\eeq{\end{equation}}
\def\bea{\begin{eqnarray}}
\def\eea{\end{eqnarray}}

\newcommand{\gsim}{\lower.7ex\hbox{$\;\stackrel{\textstyle>}{\sim}\;$}}
\newcommand{\lsim}{\lower.7ex\hbox{$\;\stackrel{\textstyle<}{\sim}\;$}}

\def\<{\left\langle}
\def\>{\right\rangle}

\begin{document}

\bibliographystyle{OurBibTeX}

\begin{titlepage}

\vspace*{-15mm}
\begin{flushright}
MPP-2007-121\\
SHEP-07-30
\end{flushright}
\vspace*{3mm}

\begin{center}
{
\sffamily\LARGE
Lepton Flavour Violation in the Constrained MSSM \\[2mm]
with Constrained Sequential Dominance
}
\\[8mm]
S.~Antusch$^{\star}$
\footnote{E-mail: \texttt{antusch@mppmu.mpg.de}}, 
S.~F.~King$^{\dagger}$
\footnote{E-mail: \texttt{sfk@hep.phys.soton.ac.uk}}
\\[1mm]

\end{center}
\vspace*{0.50cm}
\centerline{$^{\star}$ \it 
Max-Planck-Institut f\"ur Physik (Werner-Heisenberg-Institut),}
\centerline{\it 
F\"ohringer Ring 6, D-80805 M\"unchen, Germany}
\vspace*{0.1cm}
\centerline{$^\dagger$ \it School of Physics and Astronomy,}
\centerline{\it  
University of Southampton,
Southampton, SO17 1BJ, UK}
\vspace*{0.50cm}

\begin{abstract}

\noindent 
We consider charged Lepton Flavour Violation (LFV) in
the Constrained Minimal Supersymmetric Standard Model,
extended to include the see-saw mechanism with 
Constrained Sequential Dominance (CSD), where CSD provides 
a natural see-saw explanation of tri-bimaximal neutrino mixing. 
When charged lepton corrections
to tri-bimaximal neutrino mixing are included, we discover 
characteristic correlations among the LFV
branching ratios, depending on the mass ordering of the
right-handed neutrinos, with 
a pronounced dependence on the leptonic mixing 
angle $\theta_{13}$ (and in some cases also on the Dirac 
CP phase $\delta$).
\end{abstract}

\end{titlepage}

\newpage

\setcounter{footnote}{0}

\section{Introduction}
Over the past decade neutrino physics has revealed the surprising 
fact not only that neutrinos have mass, but also 
that lepton mixing must involve two large mixing angles,
commonly referred to as the atmospheric angle 
$\theta_{23}$ and the solar angle $\theta_{12}$
\cite{King:2003jb}.
The latest neutrino oscillation data 
\cite{Strumia:2006db} is consistent with 
tri-bimaximal lepton mixing \cite{Harrison:2002er}.
Theoretical attempts to reproduce
this structure typically produce tri-bimaximal mixing
in the neutrino sector \cite{deMedeirosVarzielas:2005ax}, 
with charged lepton mixing
giving important corrections to the physical lepton mixing.
For example, in the see-saw mechanism \cite{seesaw},  
sequential dominance (SD) \cite{King:1998jw}
is well known to provide a natural explanation of hierarchical
neutrino mass together with large neutrino mixing angles.
When certain constraints are imposed on the neutrino Yukawa matrix elements
then tri-bimaximal neutrino mixing can result from such a
constrained sequential dominance (CSD) \cite{King:2005bj}.
Charged lepton corrections can provide calculable deviations 
from tri-bimaximal mixing, resulting in predictive neutrino 
mixing sum rules \cite{King:2005bj,Masina:2005hf,Antusch:2005kw}
which may be proved with future long baseline neutrino experiments
\cite{Antusch:2007rk}.

When neutrino mass models are combined with supersymmetry 
(SUSY) then lepton flavour violation (LFV) is an inevitable
consequence \cite{Borzumati:1986qx,Hisano:1995cp,King:1998nv}.
In the constrained minimal supersymmetric standard model
(CMSSM), in which the soft scalar mass matrices are
described by a single universal soft high energy parameter
$m_0$, and a universal trilinear parameter $A_0$,
then the only source of LFV is due to RGE running effects,
and in this case the connection between LFV processes and 
neutrino mass models has received a good deal of attention \cite{everyone}.
In the case of SD models it has been shown that LFV could reveal
direct information about the neutrino Yukawa couplings 
in the diagonal charged lepton basis, depending on the
particular nature of the SD, for example whether the dominant
right-handed neutrino is the heaviest one or the lightest one
\cite{Blazek:2001zm,Blazek:2002wq}.
For example if the dominant right-handed neutrino is the
heaviest one, then large rates for $\tau \rightarrow \mu \gamma$
are expected \cite{Blazek:2001zm,Blazek:2002wq}.
However even in this case, the amount of information one
can deduce is limited due to the large number of
unconstrained Yukawa couplings. 

In this paper we consider LFV for the case of CSD, where the
number of independent neutrino Yukawa couplings is reduced. 
In this case the LFV predictions are also sensitive to the
charged lepton mixings, so some further assumptions
are required in order to make predictions.
In addition to tri-bimaximal mixing via CSD,
we shall also additionally assume CKM-like
charged lepton corrections. This will lead to interesting
correlations in LFV muon and tau decays, independent
of the SUSY mass parameters, and Yukawa couplings,
providing quite specific predictions for LFV.

\section{(Constrained) Sequential Dominance}
Sequential Dominance (SD) \cite{King:1998jw} represents classes of neutrino models where large 
lepton mixing angles and small hierarchical neutrino masses can be readily explained
within the see-saw mechanism. 
To understand how Sequential Dominance works, we begin by
writing the right-handed neutrino Majorana mass matrix $M_{\mathrm{RR}}$ in
a diagonal basis as 
\begin{equation}
M_{\mathrm{RR}}=
\begin{pmatrix}
M_A & 0 & 0 \\
0 & M_B & 0 \\
0 & 0 & M_C%
\end{pmatrix}.
\label{MRR}
\end{equation}
We furthermore write the neutrino (Dirac) Yukawa matrix $\lambda_{\nu}$ in
terms of $(1,3)$ column vectors $A_i,$ $B_i,$ $C_i$ as
\begin{equation}\label{Eq:YnuCSD}
Y_{\nu }=
\begin{pmatrix}
A & B & C
\end{pmatrix},
  \label{Yukawa}
\end{equation} 
using left-right convention.
The term for the light neutrino masses in the effective Lagrangian (after electroweak symmetry breaking), resulting from integrating out the massive right
handed neutrinos, is
\begin{equation}
\mathcal{L}^\nu_{eff} = \frac{(\nu_{i}^{T} A_{i})(A^{T}_{j} \nu_{j})}{M_A}+\frac{(\nu_{i}^{T} B_{i})(B^{T}_{j} \nu_{j})}{M_B}
+\frac{(\nu_{i}^{T} C_{i})(C^{T}_{j} \nu_{j})}{M_C}  \label{leff}
\end{equation}
where $\nu _{i}$ ($i=1,2,3$) are the left-handed neutrino fields.
Sequential dominance then corresponds to the third
term being negligible, the second term subdominant and the first term
dominant:
\begin{equation}\label{SDcond}
\frac{A_{i}A_{j}}{M_A} \gg
\frac{B_{i}B_{j}}{M_B} \gg
\frac{C_{i}C_{j}}{M_C} \, .
\label{SD1}
\end{equation}
In addition, we shall shortly see that small $\theta_{13}$ 
and almost maximal $\theta_{23}$ require that 
\begin{equation}
|A_1|\ll |A_2|\approx |A_3|.
\label{SD2}
\end{equation}
Without loss of generality, then, we shall label the dominant
right-handed neutrino and Yukawa couplings as $A$, the subdominant
ones as $B$, and the almost decoupled (sub-subdominant) ones as $C$. 
Note that the mass ordering of right-handed neutrinos is 
not yet specified. Again without loss of generality we shall 
order the right-handed neutrino masses as $M_1<M_2<M_3$,
and subsequently identify $M_A,M_B,M_C$ with $M_1,M_2,M_3$
in all possible ways. LFV in some of these classes of SD models has 
been analysed in \cite{Blazek:2002wq}.

Writing 
$A_\alpha = |A_\alpha| e^{i \phi_{A_\alpha}}$, 
$B_\alpha = |B_\alpha| e^{i \phi_{B_\alpha}}$, 
$C_\alpha = |C_\alpha| e^{i \phi_{C_\alpha}}$ and working in the mass basis
of the charged leptons, under the SD condition Eq.~(\ref{SD1}),
we obtain for the neutrino mixing angles 
\cite{King:1998jw}:
\begin{subequations}\label{anglesSD}\begin{eqnarray}
\label{Eq:t23} 
\tan \theta^{\nu}_{23} &\approx& \frac{|A_2|}{|A_3|}\;, \\
\label{Eq:t12}
\tan \theta^{\nu}_{12} &\approx& 
\frac{|B_1|}{c_{23}|B_2|\cos \tilde{\phi}_2 - 
s_{23}|B_3|\cos \tilde{\phi}_3  } \;,\\
\label{Eq:t13}
\theta^{\nu}_{13} &\approx& 
e^{i (\tilde{\phi} + \phi_{B_1} - \phi_{A_2})}
\frac{|B_1| (A_2^*B_2 + A_3^*B_3) }{\left[|A_2|^2 + |A_3|^2\right]^{3/2} }
\frac{M_A}{M_B} 
+\frac{e^{i (\tilde{\phi} + \phi_{A_1} - \phi_{A_2})} |A_1|}
{\sqrt{|A_2|^2 + |A_3|^2}} ,
\end{eqnarray}\end{subequations}
and for the masses
\begin{subequations}\label{massesSD}\begin{eqnarray}
\label{Eq:m3} m_3 &\approx& \frac{(|A_2|^2 + |A_3|^2)v^2}{M_A}\;, \\
\label{Eq:m2} m_2 &\approx& \frac{|B_1|^2 v^2}{s^2_{12} M_B}\;, \\
\label{Eq:m1}m_1 &\approx& {\cal O}(|C|^2 v^2/M_C) \;. 
\end{eqnarray}\end{subequations}
As in \cite{King:1998jw}
the PMNS phase $\delta$ is fixed by the requirement that we have already 
imposed in Eq.~(\ref{Eq:t12}) that $\tan(\theta_{12})$ is real and positive,
\begin{eqnarray}
\lefteqn{ \label{real12} c_{23}|B_2|\sin \tilde{\phi}_2 \;\approx\; s_{23}|B_3|\sin \tilde{\phi}_3 \; ,}  \\
\lefteqn{\label{pos12} c_{23}|B_2|\cos \tilde{\phi}_2 - s_{23}|B_3|\cos \tilde{\phi}_3 \;>\; 0\; ,}
\end{eqnarray} 
where
\bea
\tilde{\phi}_2 & \equiv & \phi_{B_2}-\phi_{B_1}-\tilde{\phi}+\delta\; ,
\nonumber \\
\tilde{\phi}_3 & \equiv & \phi_{B_3}-\phi_{B_1}+\phi_{A_2}-\phi_{A_3}
-\tilde{\phi}+\delta \; .
\label{tildephi23}
\eea
The phase $\tilde{\phi}$ is fixed by the requirement (not yet imposed
in Eq.~(\ref{Eq:t13}))
that the angle $\theta_{13}$ is real and positive.
In general this condition is rather complicated since the expression
for $\theta_{13}$ is a sum of two terms.
However if, for example, $A_1=0$ then $\tilde{\phi}$ is fixed by:
\beq
\tilde{\phi}\approx \phi_{A_2}-\phi_{B_1}-\zeta
\label{tildephi}
\eeq
where 
\beq
\zeta = \arg\left(A_2^*B_2 + A_3^*B_3  \right).
\label{eta}
\eeq
Eq.~(\ref{eta}) may be expressed as 
\beq
\tan \zeta \approx \frac{|B_2|s_{23}s_2+|B_3|c_{23}s_3}
{|B_2|s_{23}c_2+|B_3|c_{23}c_3}\,.
\label{taneta}
\eeq
Inserting $\tilde{\phi}$ of Eq.~(\ref{tildephi}) into 
Eqs.~(\ref{real12}), (\ref{tildephi23}), we obtain a relation
which can be expressed as
\beq
\tan (\zeta +\delta) \approx \frac{|B_2|c_{23}s_2-|B_3|s_{23}s_3}
{-|B_2|c_{23}c_2+|B_3|s_{23}c_3}\,.
\label{tanetadelta}
\eeq
In Eqs.~(\ref{taneta}), (\ref{tanetadelta}) we have written
$s_i=\sin \zeta_i$, $c_i=\cos \zeta_i$, where we have defined
\beq
\zeta_2\equiv \phi_{B_2}-\phi_{A_2}\;, \ \ \zeta_3\equiv
\phi_{B_3}-\phi_{A_3}\;,
\label{eta23}
\eeq
which are invariant under a charged lepton phase transformation.
The reason why the see-saw parameters only involve two invariant
phases rather than the usual six, is due to the SD assumption
in Eq.~(\ref{SD1})
that has the effect of effectively decoupling the right-handed neutrino 
of mass $M_C$ from the see-saw mechanism, which removes three phases,
together with the further assumption (in this case) of
$A_1=0$, which removes another phase.

\subsection{CSD and tri-bimaximal neutrino mixing}
Tri-bimaximal neutrino mixing \cite{Harrison:2002er} corresponds to the choice
\cite{King:2005bj}:
\begin{eqnarray}
|A_{1}| &=&0,  \label{tribicondsd} \\
\text{\ }|A_{2}| &=&|A_{3}|,  \label{tribicondse} \\
|B_{1}| &=&|B_{2}|=|B_{3}|,  \label{tribicondsa} \\
A^{\dagger }B &=&0.  \label{zero}
\end{eqnarray} 
This is called constrained sequential dominance (CSD)
\cite{King:2005bj}.
For example, a neutrino Yukawa matrix
in the notation of Eq.~(\ref{Eq:YnuCSD}), which satisfies 
the CSD conditions in Eqs.~(\ref{tribicondsd})-(\ref{zero}),
may be taken to be:
\begin{equation}
Y_{\nu }=
\begin{pmatrix}
0 & be^{i \beta_2} & c_1\\
-ae^{i \beta_3} & be^{i \beta_2} & c_2\\
ae^{i \beta_3} & be^{i \beta_2} & c_3
\end{pmatrix},
\label{Y1}
\end{equation}
where $C$ is not constrained by CSD, since it only gives a sub-subdominant
contribution to the neutrino mass matrix, so we have written
it as $C=(c_1,c_2,c_3)$ above. 
CSD leads to tri-bimaximal mixing in the neutrino mass matrix $m_\nu$, i.e.\ to
\begin{eqnarray}
V^\dagger_{\nu_\mathrm{L},\mathrm{tri}} =
\left(\begin{array}{ccc}
\!\sqrt{2/3}&1/\sqrt{3} &0\!\\
\!-1/\sqrt{6}&1/\sqrt{3}&1/\sqrt{2}\!\\
\!1/\sqrt{6}&-1/\sqrt{3}&1/\sqrt{2}\!
\end{array}
\right).
\end{eqnarray}

\section{Charged lepton corrections}
The form of the PMNS matrix will depend on the 
charged lepton Yukawa matrix whose diagonalisation will
result in a charged lepton mixing matrix $V_{e_\mathrm{L}}$
which must be combined with $V^\dagger_{\nu_\mathrm{L}}$
to form $U_{\mathrm{PMNS}}$. The resulting lepton
mixing matrix will therefore not be precisely of the tri-bimaximal
form, even in theories that predict precise tri-bimaximal
neutrino mixing.
We consider here the case that CSD holds in a basis where the charged
lepton mass matrix is not exactly diagonal, but corresponds to small
mixing. This is a situation, often encountered in realistic models
\cite{King:2005bj,Antusch:2005kw,reviews}.  

In the presence of charged lepton corrections, the prediction 
of tri-bimaximal neutrino mixing is not
directly experimentally accessible. However, this challenge can be
overcome when we make the additional assumption that the charged
lepton mixing matrix has a CKM-like structure, in the sense that
$V_{e_\mathrm{L}}$ is dominated by a 1-2 mixing
$\theta \equiv \theta^e_{12}$, i.e.\ that
its elements $(V_{e_\mathrm{L}})_{13}$, $(V_{e_\mathrm{L}})_{23}$,
$(V_{e_\mathrm{L}})_{31}$ and $(V_{e_\mathrm{L}})_{32}$ are very small
compared to $(V_{e_\mathrm{L}})_{ij}$ ($i,j = 1,2$).
In the following, we shall take these elements to be 
approximately zero, i.e.\
\begin{eqnarray}\label{Eq:AssumptionForUe}\label{Eq:Ue}
V_{e_\mathrm{L}} \approx 
P\left(\begin{array}{ccc}
\!c_\theta& -s_\theta e^{- i \lambda}&0\!\\
\!s_\theta &c_\theta e^{- i \lambda}&0\!\\
\!0&0&1\!
\end{array}
\right), 
\end{eqnarray}
where $c_\theta \equiv \cos \theta $,
$s_\theta \equiv \sin \theta $,
$\lambda$ is a phase required to diagonalise the charged lepton
mass matrix \cite{King:2005bj}, and $P$ is a diagonal matrix of phases 
$P=diag(e^{i\omega_1},e^{i\omega_2},e^{i\omega_3})$
which are chosen to remove phases from 
the product $V_{e_\mathrm{L}} V^\dagger_{\nu_\mathrm{L}}$
to yield the physical $U_{\mathrm{PMNS}}$.
In the present case it is convenient to choose
$\omega_1=0$, $\omega_2=\lambda$, $\omega_3=0$,
to yield,
\begin{eqnarray}\label{Eq:AssumptionForUe}\label{Eq:Ue}
V_{e_\mathrm{L}} \approx 
\left(\begin{array}{ccc}
\!c_\theta& -s_\theta e^{- i \lambda}&0\!\\
\!s_\theta e^{ i \lambda} &c_\theta &0\!\\
\!0&0&1\!
\end{array}
\right), 
\end{eqnarray}
With this choice, then by constructing $U_{\mathrm{PMNS}}$
and comparing to the Standard PDG form of this matrix, one
obtains, by comparing with Eq.(82) of \cite{King:2005bj}, 
\beq
\lambda = \delta - \pi
\eeq
where $\delta$ is the Standard PDG CP violating oscillation phase.
Also note that $\lambda \approx \delta_{22}-\delta_{12}$
where $\delta_{ij}= \arg M^{\mathrm{e}}_{ij}$.

We remark that the assumption that the charged lepton mixing
angles are dominated by $(1,2)$ Cabibbo-like mixing arises
in many generic classes of flavour models in
the context of unified theories of fundamental interactions, where
quarks and leptons are joined in representations of the unified gauge
symmetries~\cite{King:2005bj,Antusch:2005kw,reviews}. 
Under this assumption, it follows directly from
Eq.~(\ref{Eq:PMNS_Definition}) that $(U_{\mathrm{PMNS}})_{31}$,
$(U_{\mathrm{PMNS}})_{32}$ and $(U_{\mathrm{PMNS}})_{33}$ are
independent of $V_{e_\mathrm{L}}$, and depend only on the
diagonalisation matrix $V^\dagger_{\nu_\mathrm{L}}$ of the neutrino
mass matrix. This leads to the parameterization-independent relations \cite{Antusch:2007rk}:
\begin{subequations}\label{Eq:InvSumrules}\begin{eqnarray}
\label{Eq:InvSumrule1} |(V^\dagger_{\nu_\mathrm{L}})_{31}| &\approx& 
  |(U_{\mathrm{PMNS}})_{31}|\;,\\
\label{Eq:InvSumrule2} |(V^\dagger_{\nu_\mathrm{L}})_{32}| &\approx& 
  |(U_{\mathrm{PMNS}})_{32}|\;,\\
\label{Eq:InvSumrule3} |(V^\dagger_{\nu_\mathrm{L}})_{33}| &\approx&
  |(U_{\mathrm{PMNS}})_{33}|\;.
\end{eqnarray}\end{subequations} 
In addition to the assumption that $V_{e_\mathrm{L}}$ is of the form
of Eq.~(\ref{Eq:AssumptionForUe}) for tri-bimaximal neutrino mixing the 1-3
mixing in the neutrino mass matrix is zero,
\begin{equation}\label{eq:th13nu}
(V^\dagger_{\nu_\mathrm{L}})_{13} = 0 \,.
\end{equation}
Using Eq.~(\ref{eq:th13nu}) and applying the 
standard PDG parameterization of the PMNS matrix (see e.g.\ \cite{PDG}), 
Eq.~(\ref{Eq:InvSumrule1}) leads to the sum rule \cite{King:2005bj,Masina:2005hf,Antusch:2005kw}:
\begin{equation}\label{eq:sumrule_abs}
s_{23}^\nu s_{12}^\nu \approx
\left| s_{23}s_{12} - s_{13}c_{23}c_{12}e^{\I\delta} \right|
\approx 
s_{23}s_{12} - s_{13}c_{23}c_{12} \cos(\delta) \,,
\end{equation}
where the last step holds to leading order in $s_{13}$. This sum rule can be used to test tri-bimaximal
($\theta^{\nu}_{12} = \arcsin (\tfrac{1}{\sqrt{3}})$) structure of the
neutrino mass matrix in the presence of CKM-like charged lepton corrections.

\section{LFV in CSD with charged lepton corrections}\label{Sec:FlavourLGinSD}

When dealing with LFV it is convenient to work in the 
basis where the charged lepton mass matrix is diagonal.
Let us now discuss the consequences of charged lepton corrections of
the form of Eq.~(\ref{Eq:Ue}) for the neutrino Yukawa matrix with CSD.
After re-diagonalising the charged lepton mass matrix,
resulting in the assumed charged lepton mixing 
matrix in Eq.~(\ref{Eq:Ue}), 
$Y_\nu$ in Eq.~(\ref{Y1}) becomes transformed as:
\begin{eqnarray}
Y_\nu \rightarrow Y'_\nu = V_{e_L} \,Y_\nu \;.
\end{eqnarray}
In the diagonal charged lepton mass basis the neutrino Yukawa
matrix therefore becomes:
\begin{equation}
Y'_{\nu }=\begin{pmatrix}
A' & B' & C'
\end{pmatrix}
=\begin{pmatrix}
a \,s_\theta e^{- i \lambda}e^{i \beta_3} & 
b  \,(c_\theta - s_\theta e^{- i \lambda} )e^{i \beta_2} &
(c_1c_\theta-c_2 s_\theta e^{- i \lambda})\\
-a \,c_\theta e^{i \beta_3} & 
b \,(c_\theta + s_\theta e^{i \lambda})e^{i \beta_2} &  (c_1 s_\theta e^{i
  \lambda} +c_2 c_\theta )\\
ae^{i \beta_3} & 
be^{i \beta_2} & c_3
\end{pmatrix}.
\label{Y2}
\end{equation}
where the column vectors $A',B',C'$ are now defined in the
diagonal charged lepton basis according to Eq.~(\ref{Y2}).
Thus the results in Eqs.~(\ref{anglesSD}) with the redefined 
column vectors $A',B',C'$ now yield the physical lepton mixing angles
since these are equal to the neutrino mixing angles in the diagonal
charged lepton basis of Eq.~(\ref{Y2}).

At leading order in a mass insertion (MI) approximation 
\cite{Borzumati:1986qx,Hisano:1995cp}
the branching fractions of LFV processes are given by
\beq
Br_{ij} \equiv {\rm Br}(l_i \rightarrow l_j \gamma) \approx 
        \frac{\alpha^3}{G_F^2}
        f(M_2,\mu,m_{\tilde{\nu}}) 
        |m_{\tilde{L}_{ij}}^2|^2 \xi_{ij}  \tan ^2 \beta
    \label{eq:BR(li_to_lj)} \; ,
\eeq
where $l_1=e, l_2=\mu , l_3=\tau$,
and where the off-diagonal slepton doublet mass squared is given 
in the leading log approximation (LLA) of the CMSSM by
\beq
m_{\tilde{L}_{ij}}^{2(LLA)}
\approx -\frac{(3m_0^2+A_0^2)}{8\pi ^2}K_{ij} \; ,
\label{lla}
\eeq
with the leading log coefficients given by
\bea
K_{21} & = & A'_2{A'_1}^*\ln \frac{\Lambda}{M_A} + 
B'_2{B'_1}^*\ln \frac{\Lambda}{M_B} + 
C'_2{C'_1}^*\ln \frac{\Lambda}{M_C} \; ,\nonumber \\
K_{32} & = & A'_3{A'_2}^*\ln \frac{\Lambda}{M_A} + 
B'_3{B'_2}^*\ln \frac{\Lambda}{M_B} + 
C'_3{C'_2}^*\ln \frac{\Lambda}{M_C} \; ,\nonumber \\
K_{31} & = & A'_3{A'_1}^*\ln \frac{\Lambda}{M_A} + 
B'_3{B'_1}^*\ln \frac{\Lambda}{M_B} + 
C'_3{C'_1}^*\ln \frac{\Lambda}{M_C}\; .
\label{Cij}
\eea
The factors $\xi_{ij}$ in Eq.~(\ref{eq:BR(li_to_lj)}) 
represent the ratio of the leptonic partial
width to the total width, 
\beq
\xi_{ij}=\frac{\Gamma (l_i\rightarrow l_j\nu_i \overline{\nu}_j)}
{\Gamma (l_i \rightarrow {\rm all})}\; .
\eeq
Clearly $\xi_{21}=1$ but $\xi_{32}$ is non-zero and must be included
for correct comparison with the experimental limit on the branching
ratio for $\tau \rightarrow \mu \gamma$. This factor is frequently forgotten
in the theoretical literature.

If LFV is only induced by RG effects from $Y'_\nu$ on the soft breaking
terms, as in the CMSSM, then in the LLog and MI approximation, the
branching ratios for LFV charged lepton decays, like $\ell_i \to
\ell_j \gamma$, are proportional to
\begin{equation}
\!\!\!\!\!\!\!\! 
Br_{ij} \propto |K_{ij}|^2=
|(A' A^{' \dagger})_{ij} \ln (\Lambda/M_A) +
(B' B^{' \dagger})_{ij} \ln (\Lambda/M_B) +
(C' C^{' \dagger})_{ij} \ln (\Lambda/M_C)|^2  \!.
\label{props}
\end{equation}
We have only assumed so far that the right-handed neutrino mass matrix has the
diagonal form shown in Eq.~(\ref{MRR}), $M_{RR}= \mbox{diag}(M_A,M_B,M_C)$
with the dominant right-handed
neutrino labelled by $A$, the leading subdominant one labelled by $B$,
and the decoupled one labelled by $C$. However
the masses of the right-handed neutrinos are not yet ordered, 
and we have not yet specified which one is the lightest and so on.
After ordering
$M_A,M_B,M_C$ according to their size, there are six possible forms of
$Y'_\nu$ obtained from permuting the columns, with the convention 
always being that the dominant one is labelled by A, and so on.
In particular the third column of the neutrino Yukawa matrix
could be $A'$, $B'$ or $C'$ depending on which of 
$M_A$, $M_B$ or $M_C$ is the heaviest.

In hierarchical models, the $(3,3)$ elements of the Yukawa matrices
describing quarks and charged leptons
are amongst the largest elements in the Yukawa matrices. In unified
models this will also be the case for the neutrino Yukawa matrix.
If the heaviest right-handed neutrino mass is
$M_A$ then the third column of the neutrino Yukawa matrix 
will consist of the $A'$ column, and since $Y^{\nu}_{33}=A'_3$
and $A'_1 \sim A'_2 \sim A'_3 \sim a$ then
we conclude that all elements of $A'$ must dominate over those
of $B',C'$, and hence all LFV processes will be determined approximately by
$(A' A^{' \dagger})_{ij}$. Similarly 
if the heaviest right-handed neutrino mass is
$M_B$ then the third column of the neutrino Yukawa matrix 
will consist of the $B'$ column, and since $Y^{\nu}_{33}=B'_3$
and $B'_1 \sim B'_2 \sim B'_3 \sim b$ then
we conclude that all elements of $B'$ must dominate over those
of $A',C'$, and hence all LFV processes will be determined approximately by
$(B' B^{' \dagger})_{ij}$. Finally 
if the heaviest right-handed neutrino mass is
$M_C$ then the third column of the neutrino Yukawa matrix 
will consist of the $C'$ column which contains the large
element $Y^{\nu}_{33}=C'_3$. However in this case 
we cannot conclude that all elements of $C'$ must dominate over those
of $A',B'$ for the determination of LFV processes since 
the elements $c_1, c_2$ are undetermined by the see-saw
mechanism and could even be set equal to zero.
Nevertheless it is possible that in this case
all elements of $C'$ could dominate over those
of $A',B'$ and hence all LFV processes could be determined approximately by
$(C' C^{' \dagger})_{ij}$.
In the following we consider the LFV predictions arising from 
the three cases 
\begin{eqnarray}
M_3 = M_A \;,\quad M_3 = M_B \;,\quad M_3 = M_C \;,
\end{eqnarray} 
corresponding to the dominant Yukawa
columns being $A'$, $B'$, $C'$, respectively.

\section{Predictions for the ratios of LFV branching ratios}
After ordering
$M_A,M_B,M_C$ according to their size, there are six possible forms of
$Y'_\nu$ obtained from permuting the columns, with the convention 
always being that the dominant one is labeled by $A'$, and so on.
In particular the third column of the neutrino Yukawa matrix
could be $A'$, $B'$ or $C'$ depending on which of 
$M_A$, $M_B$ or $M_C$ is the heaviest.
If the heaviest right-handed neutrino mass is
$M_A$ then the third column of the neutrino Yukawa matrix 
will consist of the (re-ordered) first column of Eq.~(\ref{Y2})
and assuming $Y^{' \nu}_{33}\sim 1$
we conclude that all LFV processes will be determined approximately by
the first column of Eq.~(\ref{Y2}). Similarly 
if the heaviest right-handed neutrino mass is
$M_B$ then we conclude that  
all LFV processes will be determined approximately by
the second column of Eq.~(\ref{Y2}).
Note that in both cases the ratios of branching ratios
are independent of the unknown Yukawa couplings which cancel,
and only depend on the charged lepton angle $\theta \equiv \theta^e_{12}$ 
(and in some cases on $\lambda$),
which in the case of tri-bimaximal neutrino mixing is related to 
the physical reactor angle by 
$\theta_{13} = \theta^e_{12}/\sqrt{2} \equiv \theta /\sqrt{2}$
\cite{King:2005bj,Antusch:2005kw}. Also note that
$\lambda = \delta - \pi$
where $\delta$ is the Standard PDG CP violating oscillation phase.
The predictions for these two cases will now be discussed
in detail. We will also comment on the third case $M_3 = M_C$, which is less predictive, 
and give an explicit minimal example.

\subsection{$\boldsymbol{M_3 = M_A}$}
\begin{figure}[t]
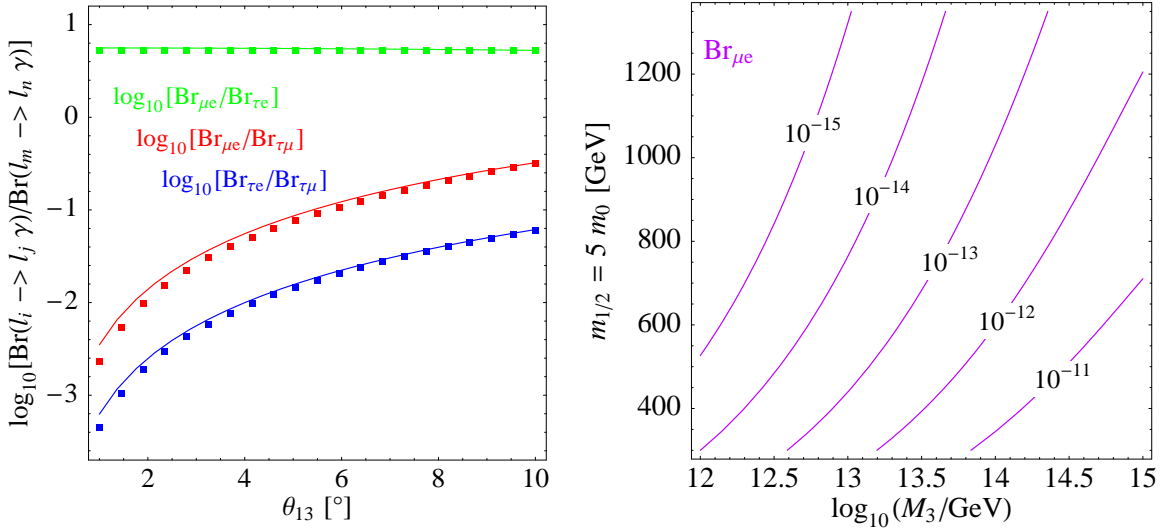

\centering
$\ensuremath{\vcenter{\hbox{\includegraphics[scale=0.85]{plotM3_MA.eps}}}}$ \;\;
$\ensuremath{\vcenter{\hbox{\includegraphics[scale=0.9]{plotM3_MA_m0.eps}}}}$ 
 \caption{\label{fig:CSDplotsMA}
The left panel shows the ratios of branching ratios $Br_{ij}$ of LFV processes 
$\ell_i \to \ell_j \gamma$ in CSD for $M_3 = M_A$ 
with right-handed neutrino masses $M_1=10^8$ GeV, $M_2=5\times 10^8$
GeV and $M_3=10^{14}$ GeV. Here the solid lines show the (naive)
prediction, from the MI and LLog approximation and with RG running
effects for the other parameters neglected, while the dots show the explicit numerical
computation (using SPheno2.2.2.~\cite{Porod:2003um} extended by software
packages for LFV branching ratios and neutrino mass matrix running \cite{Arganda:2005ji,Antusch:2006vw}) 
with universal CMSSM parameters chosen 
as $m_0=750$ GeV, $m_{1/2}=750$ GeV, $A_0 = 0$ GeV, $\tan \beta = 10$ 
and $\mbox{sign}(\mu)=+1$. 
The right panel shows the predictions (from full computation) for
$Br_{\mu e} = Br (\mu \to e \gamma)$ in the CMSSM extended by the 
see-saw mechanism with CSD for the case $M_3 = M_A$ 
with $\theta_{13} = 3^\circ$ and $\delta = 0$. In this panel
we have chosen the CMSSM parameters to satisfy 
$A_0 = 0$ GeV, $\tan \beta = 10$ and $m_{1/2}=5 m_0$, which approximately 
corresponds to the successful stau co-annihilation region of LSP neutralino 
dark matter (DM) giving $\Omega_{DM}$ within the current WMAP limits.} 
\end{figure}

In this case, assuming that the third column of the neutrino
Yukawa matrix (associated with the heaviest right-handed
neutrino and hence the largest Yukawa couplings)
is the dominant column $A'$ associated with the
atmospheric neutrino of mass $m_3$,
one can read off from Eq.~(\ref{props}) and Eq.~(\ref{Y2})
that the $Br_{ij} \equiv Br(\ell_i \to \ell_j \gamma)$ now satisfy
\begin{eqnarray}
Br_{\mu e} &\propto& |a^2 \,s_\theta c_\theta|^2
\,\xi_{\mu e} \;,\\ 
Br_{\tau e} &\propto& |a^2
\,s_\theta |^2 \,\xi_{\tau e} \;,\\ 
Br_{\tau \mu}
&\propto& |a^2|^2 \,\xi_{\tau \mu} \;.
\end{eqnarray}
Note that $\theta \equiv \theta^e_{12} = \sqrt{2}\theta_{13}$,
so there is a direct (and simple) connection to the 
measurable lepton mixing angle $\theta_{13}$ in neutrino oscillation experiments in this case. In particular, we predict
\begin{eqnarray}
\frac{Br_{\mu e}}{Br_{\tau \mu}} &=&  (s_\theta c_\theta)^2  \frac{\xi_{\mu e}}{\xi_{\tau \mu}} 
\;\; = \;\; \left[\tfrac{1}{2} \sin(2 \sqrt{2} \theta_{13})\right]^2 \frac{\xi_{\mu e}}{\xi_{\tau \mu}} \;,\\
\frac{Br_{\mu e}}{Br_{\tau e}} &=&  
(c_\theta)^2 \frac{\xi_{\mu e}}{\xi_{\tau e}}\;\; = \;\; \left[\cos(\sqrt{2}\theta_{13})\right]^2 
\frac{\xi_{\mu e}}{\xi_{\tau e}}\;,
\\
\frac{Br_{\tau e}}{Br_{\tau \mu}} &=&  (s_\theta)^2 \;\; = \;\; \left[ \sin(\sqrt{2}\theta_{13})\right]^2.
\end{eqnarray}
The predictions for the ratios of branching ratios as a function of $\theta_{13}$ as well as for 
$Br_{\mu e}$, for some sample choice of parameters, are
shown in Fig.~\ref{fig:CSDplotsMA}.

\subsection{$\boldsymbol{M_3 = M_B}$}
\begin{figure}[t]
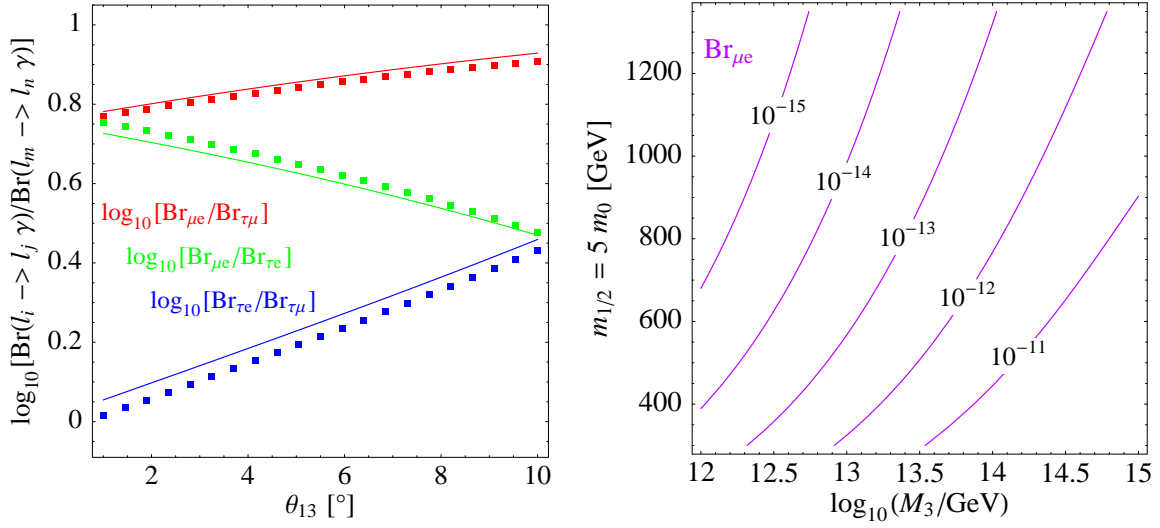

\centering
$\ensuremath{\vcenter{\hbox{\includegraphics[scale=0.85]{plotM3_MB.eps}}}}$ \;\;
$\ensuremath{\vcenter{\hbox{\includegraphics[scale=0.89]{plotM3_MB_m0.eps}}}}$ 
 \caption{\label{fig:CSDplotsMB} 
The left panel shows the ratios of branching ratios $Br_{ij}$ of LFV processes 
$\ell_i \to \ell_j \gamma$ in CSD for $M_3 = M_B$ and for $\delta = 0$. The other parameters are chosen 
as in Fig.~\ref{fig:CSDplotsMA}. The solid lines in the left panel show the (naive)
prediction, from the MI and LLog approximation and with RG running
effects neglected, while the dots show the explicit numerical
computation. 
The right panel shows the predictions (from full computation)  for
$Br_{\mu e} = Br (\mu \to e \gamma)$ in the CMSSM extended by the 
see-saw mechanism with CSD for the case $M_3 = M_B$ 
with $\theta_{13} = 3^\circ$ and $\delta = 0$.
The other parameters are chosen 
as in Fig.~\ref{fig:CSDplotsMA}.}
\end{figure}

In this case, assuming that the third column of the neutrino
Yukawa matrix (associated with the heaviest right-handed
neutrino and hence the largest Yukawa couplings)
is the leading subdominant column $B'$ associated with the
solar neutrino of mass $m_2$,
one can read off from Eq.~(\ref{props}) and Eq.~(\ref{Y2})
that the $Br_{ij} \equiv Br(\ell_i \to \ell_j \gamma)$ now satisfy
\begin{eqnarray}
Br_{\mu e} &\propto& 
|b^2 (c_\theta -s_\theta e^{- i \lambda} ) (c_\theta + s_\theta e^{i \lambda})|^2 \,
\xi_{\mu e} \;,\\
Br_{\tau e} &\propto& |b^2 (c_\theta - s_\theta e^{- i \lambda})  |^2 \,
\xi_{\tau e} \;,\\
Br_{\tau \mu} &\propto& |b^2 (c_\theta + s_\theta e^{i \lambda})|^2  \,
\xi_{\tau \mu} \;.
\end{eqnarray}
Since $\theta \equiv \theta^e_{12} = \sqrt{2}\theta_{13}$, 
there is again a connection to the measurable lepton mixing angle 
$\theta_{13}$ in neutrino oscillation experiments. Furthermore, the branching ratios also depend on 
the phase $\lambda$, which is related to the Standard PDG CP violating oscillation phase $\delta$ by
$\lambda = \delta - \pi$. The ratios of branching ratios are predicted as 
\begin{eqnarray}
\label{Eq:M3MB_1}\frac{Br_{\mu e}}{Br_{\tau \mu}} &=&  
|c_\theta -s_\theta e^{- i \lambda} |^2  \frac{\xi_{\mu e}}{\xi_{\tau \mu}} 
\;\;=\;\; |\cos(\sqrt{2}\theta_{13}) + \sin(\sqrt{2}\theta_{13}) e^{- i \delta} |^2  \frac{\xi_{\mu e}}{\xi_{\tau \mu}} 
\;,\\
\label{Eq:M3MB_2}\frac{Br_{\mu e}}{Br_{\tau e}} &=&  
|c_\theta + s_\theta e^{i \lambda}|^2 \frac{\xi_{\mu e}}{\xi_{\tau e}}\;\;=\;\; |\cos(\sqrt{2}\theta_{13}) - 
\sin(\sqrt{2}\theta_{13}) e^{ i \delta} |^2 \frac{\xi_{\mu e}}{\xi_{\tau e}}
\;,\\
\label{Eq:M3MB_3}\frac{Br_{\tau e}}{Br_{\tau \mu}} &=&  \left| 
\frac{c_\theta -s_\theta e^{- i \lambda} }{c_\theta + s_\theta e^{i \lambda}} \right|^2 \;\;=\;\;\left| 
\frac{\cos(\sqrt{2}\theta_{13}) + \sin(\sqrt{2}\theta_{13}) e^{- i \delta} }{\cos(\sqrt{2}\theta_{13}) - 
\sin(\sqrt{2}\theta_{13}) e^{ i \delta} } \right|^2
 \!.
\end{eqnarray}
Fig.~\ref{fig:CSDplotsMB} shows the predictions for the ratios of branching ratios
as a function of $\theta_{13}$, for the example $\delta = 0$, as well as 
the prediction for $Br_{\mu e}$ for some sample choice of parameters.

\subsection{$\boldsymbol{M_3 = M_C}$}
In this case, assuming that the third column of the neutrino
Yukawa matrix (associated with the heaviest right-handed
neutrino and hence the largest Yukawa couplings)
is the most subdominant column $C'$ associated with the
lightest neutrino of mass $m_1$, assuming that $c_3 \approx 1$,
one can see from Eq.~(\ref{props}) and Eq.~(\ref{Y2})
that the $Br_{ij} \equiv Br(\ell_i \to \ell_j \gamma)$
now depend on undetermined coefficients $c_1, c_2$.
Hence we cannot make definite predictions. 
Moreover, in some cases, the subdominant column of Yukawa coupling also contributes at the same
order as the dominant one. Nevertheless, charged lepton corrections
also have an impact here. Let us therefore generalize $V_{e_L}$ to
include also a small $\theta^e_{23} \ll \theta^e_{12}$.
As a minimal case, let us furthermore consider
\begin{eqnarray}
C &=& (0,0,c)^T \;.
\end{eqnarray}
This may be viewed as minimal scenario regarding LFV, since
typically (barring cancellations) the zeros are replaced by small 
entries and since, as mentioned above, the subdominant column of $Y_\nu$ 
can not in general be neglected. For a more accurate treatment of this scenario
with respect to the charged lepton corrections, the (typically) even smaller 
$\theta^e_{13} \ll \theta^e_{23}$ can be included analogously.

\begin{figure}[t]
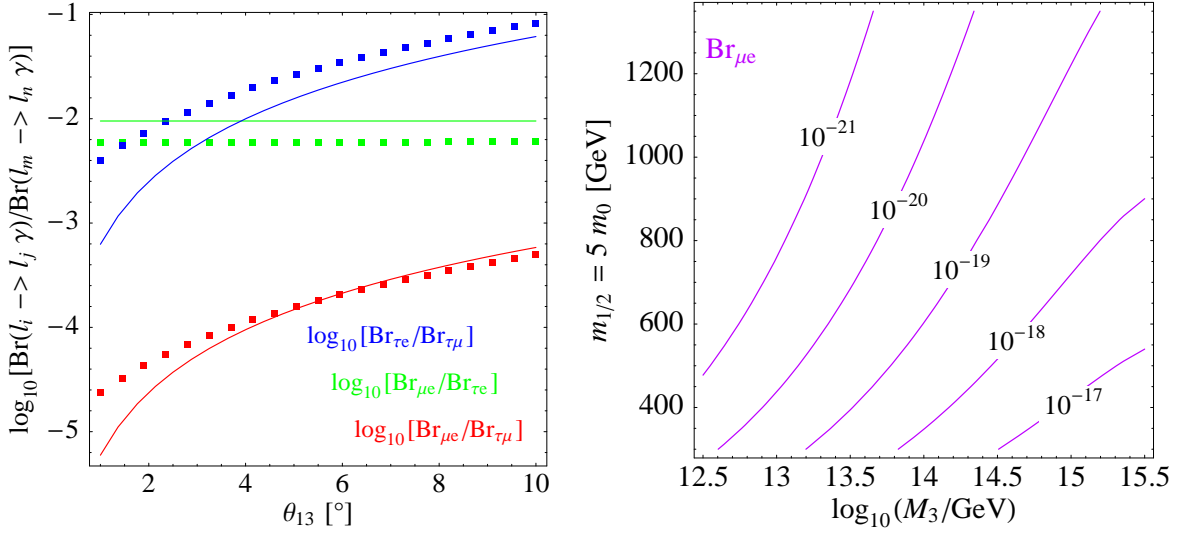

\centering
$\ensuremath{\vcenter{\hbox{\includegraphics[scale=0.85]{plotM3_MC.eps}}}}$ \;\;
$\ensuremath{\vcenter{\hbox{\includegraphics[scale=0.92]{plotM3_MC_m0.eps}}}}$ 
 \caption{\label{fig:CSDplotsMC} 
The left panel shows the ratios of branching ratios $Br_{ij}$ of LFV processes 
$\ell_i \to \ell_j \gamma$ for a minimal example with CSD and $M_3 = M_C$ described in the text. 
The solid lines in the left panel show the (naive)
prediction, from the MI and LLog approximation and with RG running
effects neglected, while the dots show the explicit numerical
computation. 
The right panel shows the predictions (from full computation) 
for $Br_{\mu e} = Br (\mu \to e \gamma)$ 
in the CMSSM extended by the 
see-saw mechanism with CSD for the case $M_3=M_C$ 
in the scenario with $m_1 = 10^{-3}$ eV and $\delta = 0$. 
The other parameters are chosen 
as in Fig.~\ref{fig:CSDplotsMA}.}
\end{figure}

Including charged lepton corrections from $\theta^e_{12}$ and $\theta^e_{23}$ 
(by $Y_\nu \to V_{e_L} Y_\nu$) leads to approximately
 \begin{eqnarray}
C' &=& (c s_{23}^e s_{12}^e,c s_{23}^e,c)^T \;,
\end{eqnarray} 
and thus to the following relations for the branching ratios:
\begin{eqnarray}
Br_{\mu e} &\propto& |c^2 (s_{23}^e)^2 s_{12}^e|^2
\,\xi_{\mu e} \;,\\ 
Br_{\tau e} &\propto& |c^2
s_{23}^e s_{12}^e|^2 \,\xi_{\tau e} \;,\\
Br_{\tau \mu} &\propto& |c^2 s_{23}^e|^2
\,\xi_{\tau \mu}\;.
\end{eqnarray}
As in the cases $M_3 = M_A$ and $M_3 = M_B$, the relation 
$\theta^e_{12}= \sqrt{2} \theta_{13}$
holds under the 
considered assumption about the charged lepton corrections. For the ratios of the branching ratios we obtain
\begin{eqnarray}
\frac{Br_{\mu e}}{Br_{\tau \mu}} &=&  
\left[ s_{12}^e s_{23}^e \right]^2  \frac{\xi_{\mu e}}{\xi_{\tau \mu}}
\;\;=\;\; \left[ \sin(\sqrt{2}\theta_{13}) s_{23}^e \right]^2  \frac{\xi_{\mu e}}{\xi_{\tau \mu}} 
\;,\\
\frac{Br_{\mu e}}{Br_{\tau e}} &=&  
\left[ s_{23}^e \right]^2 \frac{\xi_{\mu e}}{\xi_{\tau e}}
\;,\\
\frac{Br_{\tau e}}{Br_{\tau \mu}} &=&  \left[ s_{12}^e  \right]^2 
\;\;=\;\; \left[ \sin(\sqrt{2}\theta_{13}) \right]^2  
\!.
\end{eqnarray}
The predictions for the ratios of branching ratios as a function of $\theta_{13}$ as well as 
for $Br_{\mu e}$ as a function of $\theta_{13}$ and $m_{1/2}$ (set equal to $5 m_0$ as an example)  
are shown in Fig.~\ref{fig:CSDplotsMC}. To give  an explicit example, we have chosen $s_{23}^e =
\sin(\theta_{23}^{\mathrm{CKM}}) \approx 2.36^\circ$ and other parameters as stated in the caption of 
Fig.~\ref{fig:CSDplotsMA}. We would like to stress again that, in contrast to the cases $M_3 = M_A$ and $M_3 =
M_B$ discussed above, the shown results are no definite predictions for the case $M_3 = M_C$, but rather
order of magnitude examples for certain classes of models of CSD where the LFV branching ratios are
significantly smaller than for CSD with  $M_3 = M_A$ and $M_3 =M_B$. 
As can be seen from Fig.~\ref{fig:CSDplotsMC}, this scenario can be readily distinguished from the 
cases $M_3 = M_A$ and $M_3 = M_B$.

\section{Conclusions}
We have considered charged Lepton Flavour Violation (LFV) in
the Constrained Minimal Supersymmetric Standard Model,
extended to include the see-saw mechanism with 
Constrained Sequential Dominance (CSD), where CSD provides 
a natural see-saw explanation of tri-bimaximal neutrino mixing. 
When Cabibbo-like charged lepton corrections
to tri-bimaximal neutrino mixing are included, 
this leads to characteristic correlations among the LFV
branching ratios 
$Br_{\tau \mu}$, $Br_{\mu e}$ and $Br_{\tau e}$
which may be tested in future experiments.

There are two main differences between the study here 
and that in \cite{Blazek:2002wq} where predictions for LFV were
also presented for the CMSSM with SD.
The first difference is that here we have focused on the special case of CSD, 
corresponding to tri-bimaximal neutrino mixing,
where the neutrino Yukawa couplings are very tightly constrained
compared to the general SD case. The second difference is that 
we have considered the effect of charged lepton corrections,
which were not included in \cite{Blazek:2002wq}.
In particular we have mainly considered Cabibbo-like charged
lepton corrections, which when combined with CSD leads to 
a very tightly constrained scenario in which ratios of 
branching ratios depend on 
$\theta_{13}$, which is related to the charged lepton 
mixing angle $\theta^e_{12}$.
The predictions also depend crucially 
on which column of the Yukawa matrix
is associated with the heaviest right-handed neutrino $M_3$,
since this column will have the largest Yukawa couplings.

For the case $M_3=M_A$, also known as
Heavy Sequential Dominance (HSD) since the dominant
right-handed neutrino is the heaviest one, 
we find the characteristic ratios in Fig.~1.
Compared to the results in \cite{Blazek:2002wq},
the hierarchy between $Br_{\mu e}$ and $Br_{\tau \mu}$
is much milder. This can be understood from the fact that 
in \cite{Blazek:2002wq} it was assumed 
that $|A_1|\ll |A_2|\approx |A_3|\sim 1$ 
(ignoring charged lepton corrections) which led to large
$Br_{\tau \mu}$ but small $Br_{\mu e}$. 
However, including charged lepton corrections, 
we see that $|A'_1|\sim |A'_2| \sim |A'_3|\sim 1$,
leading to both large $Br_{\tau \mu}$ and large $Br_{\mu e}$.
In the present case we focus on tri-bimaximal neutrino mixing,
which before charged lepton corrections are included
implies that $|A_1|\ll |A_2|=|A_3|\sim 1$,
 corresponding to the CSD
explanation of tri-bimaximal neutrino mixing.
Then, after Cabibbo-like charged lepton corrections are included,
this leads to well defined predictions for the each of the couplings 
$|A'_1|,|A'_2|,|A'_3|$, and hence
rather precise predictions for ratios of 
$Br_{\tau \mu}$, $Br_{\mu e}$ and $Br_{\tau e}$,
which depend on $\theta_{13}$, as shown in Fig.~1.
We reemphasize that, after charged lepton corrections are
included, $|A'_1|\sim |A'_2| \sim |A'_3|\sim 1$, and hence
both $Br_{\tau \mu}$ and $Br_{\mu e}$ are large in this case,
unlike \cite{Blazek:2002wq} where charged lepton corrections were 
ignored.

In the case $M_3=M_B$, where the leading subdominant right-handed
neutrino responsible for the solar neutrino mass is the heaviest one,
the predicted ratios of branching ratios are even milder,
corresponding to the fact that all the Yukawa coupling
in this column are equal before the inclusion
of charged lepton corrections, 
$|B_1|=|B_2|=|B_3|\sim 1$, again corresponding to the CSD
explanation of tri-bimaximal neutrino mixing.
When Cabibbo-like charged lepton
corrections are included this again leads to 
characteristic predictions for ratios of 
$Br_{\tau \mu}$, $Br_{\mu e}$ and $Br_{\tau e}$,
depending on $\theta_{13}$ (shown in Fig.~2 for $\delta = 0$) 
as well as on the Dirac CP phase $\delta$, as given in 
Eqs.~(\ref{Eq:M3MB_1}) - (\ref{Eq:M3MB_3}). 

The least predictive case is $M_3=M_C$, which includes the
case where the dominant right-handed neutrino is the lightest 
one known as light sequential dominance (LSD). In this case
the generic prediction from \cite{Blazek:2002wq} was that the
$Br_{\tau \mu}$ was generally quite small, typically
of order $Br_{\mu e}$, due to the small neutrino Yukawa couplings.
In particular the neutrino Yukawa couplings of the third column
were not considered relevant due to the large mass of the associated
right-handed neutrino, which was assumed to exceed the
GUT scale from which the RGEs were run down. Then the relevant Yukawa
couplings were those from the second column, which all take similar
(small) values leading to $Br_{\tau \mu}\sim Br_{\mu e}$.
This may also be the case here, since including charged lepton
corrections will not change this result, and CSD will only
strengthen this conclusion. However in other cases, for example 
if the RGEs are run from 
the Planck scale, the third column of the 
neutrino Yukawa matrix should not
be ignored. In Fig.~3 we considered an example of this, in which 
the LFV arises solely from the third column, and the Yukawa couplings
in this column are again determined from charged lepton corrections,
assuming that $C=(0,0,c)^T$, which may be approximately true in
practice, but which is by no means guaranteed.

In summary, the results presented here once again confirm that
$Br_{\tau \mu}$, $Br_{\mu e}$ and $Br_{\tau e}$ are all expected
to be observed in the (near) future. If they are observed 
in the ratios predicted here, for some value of $\theta_{13}$,
then this may be an indication of a high energy theory
with the characteristics of the CMSSM 
extended to include the see-saw mechanism with CSD,
corresponding to tri-bimaximal neutrino mixing
corrected by Cabibbo-like charged lepton mixing angles.

\section*{Acknowledgements}
S.~A.\ would like to thank M.~J.~Herrero and E.~Arganda for many very
helpful discussions about LFV in SUSY see-saw scenarios and for providing
their software packages on LFV muon and tau decays. 

\section*{Appendix}
\appendix

\renewcommand{\thesection}{\Alph{section}}
\renewcommand{\thesubsection}{\Alph{section}.\arabic{subsection}}
\def\theequation{\Alph{section}.\arabic{equation}}
\renewcommand{\thetable}{\arabic{table}}
\renewcommand{\thefigure}{\arabic{figure}}
\setcounter{section}{0}
\setcounter{equation}{0}

\section{Conventions}\label{conventions}
In general, the mixing matrix in the lepton sector, the PMNS matrix
$U_{\mathrm{PMNS}}$, is defined as the matrix which appears in the
electroweak coupling to the $W$ bosons expressed in terms of lepton
mass eigenstates. With the mass matrices of charged leptons
$M_\mathrm{e}$ and neutrinos $m_{\nu}$ written as
\begin{eqnarray}
{\cal L}=-  \bar{e}_L M_\mathrm{e} e_R  
- \tfrac{1}{2}\bar{\nu}_L m_{\nu} \nu_\mathrm{L}^c 
+ \text{H.c.}\; ,
\end{eqnarray}
and performing the transformation from flavour to mass basis by
 \begin{eqnarray}\label{eq:DiagMe}
V_{\mathrm{e}_\mathrm{L}} \, M_{\mathrm{e}} \,
V^\dagger_{\mathrm{e}_\mathrm{R}} =
\mbox{diag}(m_e,m_\mu,m_\tau)
 , \quad
V_{\nu_\mathrm{L}} \,m_\nu\,V^T_{\nu_\mathrm{L}} =
\mbox{diag}(m_1,m_2,m_3),
\end{eqnarray}
the PMNS matrix is given by
\begin{eqnarray}\label{Eq:PMNS_Definition}
U_{\mathrm{PMNS}} 
= V_{e_\mathrm{L}} V^\dagger_{\nu_\mathrm{L}} \,.
\end{eqnarray}
Here it is assumed implicitly that unphysical phases are removed by
field redefinitions, and $U_\mathrm{PMNS}$ contains one Dirac phase
and two Majorana phases. The latter are physical only in the case of
Majorana neutrinos, for Dirac neutrinos the two Majorana phases can be
absorbed as well. 

The standard PDG parameterization of the
PMNS matrix (see e.g.\ \cite{PDG}) is: 
\begin{eqnarray}\label{Eq:StandardParametrization}
 %\hspace{-0.5cm} 
 U_{\mathrm{PMNS}} = \left(
  \begin{array}{ccc}
  c_{12}c_{13} & 
  s_{12}c_{13} & s_{13}e^{-\I\delta}\\
  -c_{23}s_{12}-s_{13}s_{23}c_{12}e^{\I\delta} &
  c_{23}c_{12}-s_{13}s_{23}s_{12}e^{\I\delta}  &
  s_{23}c_{13}\\
  s_{23}s_{12}-s_{13}c_{23}c_{12}e^{\I\delta} &
  -s_{23}c_{12}-s_{13}c_{23}s_{12}e^{\I\delta} &
  c_{23}c_{13}
  \end{array}
  \right) \, P_\mathrm{Maj} \, ,
\end{eqnarray}
which is used in most analyses of neutrino oscillation experiments.
Here $\delta$ is the so-called Dirac CP violating phase which is in
principle measurable in neutrino oscillation experiments, and 
$P_\mathrm{Maj} = \mathrm{diag}(e^{\I \tfrac{\alpha_1}{2}}, 
e^{\I\tfrac{\alpha_2}{2}}, 0)$ contains the Majorana phases $\alpha_1,
\alpha_2$.  In the following we will use this standard
parameterization (including additional phases) also for 
$V^\dagger_{\nu_\mathrm{L}}$ and denote the
corresponding mixing angles by $\theta_{ij}^\nu$, while the mixing
angles $\theta_{ij}$ without superscript refer to the PMNS matrix.

\providecommand{\bysame}{\leavevmode\hbox to3em{\hrulefill}\thinspace}

\end{document}